\documentclass[prd,aps,12pt,nofootinbib,superscriptaddress]{revtex4-1}

\usepackage{graphicx}

\begin{document}

\title{Arrow of time in dissipationless cosmology}

\author{Varun Sahni}
\affiliation{Inter-University Centre for Astronomy and Astrophysics, Post Bag 4,
Ganeshkhind, Pune 411~007, India} %

\author{Yuri Shtanov\,$^6$}\footnotetext[6]{Author to whom any correspondence should be addressed.}
\affiliation{Bogolyubov Institute for Theoretical Physics, 14-b Metrologicheskaya
St., Kiev 03680, Ukraine} %
\affiliation{Department of Physics, Taras Shevchenko National University of Kiev,
64/13 Vladimirskaya St., Kiev 01601, Ukraine} %

\author{Aleksey Toporensky}
\affiliation{Sternberg Astronomical Institute, Moscow State University, Universitetsky
Prospekt, 13, Moscow 119992, Russia} %
\affiliation{Kazan Federal University, Kremlevskaya 18, Kazan 420008, Russia} %

\begin{abstract}
It is generally believed that a cosmological arrow of time must be associated with
entropy production. Indeed, in his seminal work on cyclic cosmology, Tolman introduced a
viscous fluid in order to make successive expansion/contraction cycles larger than
previous ones, thereby generating an arrow of time. However, as we demonstrate in this
letter, the production of entropy is not the only means by which a cosmological arrow of
time may emerge. Remarkably, systems which are dissipationless may nevertheless
demonstrate a preferred direction of time provided they possess {\em attractors\/}. An
example of a system with well defined attractors is scalar-field driven cosmology. In
this case, for a wide class of potentials (especially those responsible for inflation),
the attractor equation of state during expansion can have the form $p \simeq -\rho$, and
during contraction $p \simeq \rho$. If the resulting cosmology is cyclic, then the
presence of {\em cosmological hysteresis\/}, $\oint p~dV \neq 0$ during successive
cycles, causes an arrow of time to emerge in a system which is formally dissipationless.
An important analogy is drawn between the  arrow of time in cyclic cosmology and an arrow
of time in an $N$-body system of gravitationally interacting particles. We find that,
like the $N$-body system, a cyclic universe can evolve from a single past into two
futures with {\em oppositely directed\/} arrows of time.

\medskip

E-mail: varun@iucaa.ernet.in, shtanov@bitp.kiev.ua and atopor@rambler.ru
\end{abstract}



\maketitle

\setcounter{footnote}{6}

\section{Introduction}

It is a common and widespread belief that an arrow of time is invariably associated with
some form of dissipation and entropy production which usually characterize irreversible
phenomena. While it is true that dissipative phenomena do give rise to a preferred
orientation of time, we demonstrate in this letter that entropy production is not
essential for the existence of time's arrow, and that such an arrow can appear even if
the equations describing cosmic evolution are dissipationless and therefore formally time
reversible. Examples to support our argument arise from scalar-field driven cosmology
which has relevance both for a cyclic universe (section~\ref{sec:cycle}) and inflation
(section~\ref{sec:inflation}). In the case of the cyclic universe, an arrow of time
emerges on account of the presence of hysteresis: $\oint pdV \neq 0$, evaluated for any
expansion-contraction cycle. Here, $p$ is pressure and $V$ is the volume of the universe.
In this case, the increase $\delta a_{\rm max}$ in successive expansion maxima of the
scale factor is related to the amount of hysteresis through $\delta a_{\rm max}^n = A
\oint pdV$, where $n$ and $A$ depend upon the mechanism which causes the universe to turn
around and contract in a given cycle.

We speculate that the presence of an arrow of time may be caused by the existence of
attractors which single out one type of trajectory ($p \simeq -\rho$) during expansion
and a different one ($p \simeq \rho$) during contraction. This behaviour is typical of
inflationary potentials. In section~\ref{sec:cycle}, we demonstrate that, while the
scalar-field driven cyclic cosmology with hysteresis is stable, its time reversal
possesses an instability. A similar instability with respect to time reversal also exists
in the inflationary universe, discussed in section~\ref{sec:inflation}. In
section~\ref{sec:discussion}, we also draw attention to a curious similarity between the
arrow of time in a cyclic universe and an arrow of time, recently discovered, in the
gravitational clustering of dissipationless matter.

\section{Arrow of time in Cyclic Cosmology}
\label{sec:cycle}

Cyclic cosmology with dissipationless matter is usually not thought to possess an arrow
of time. Indeed, if the Friedmann--Robertson--Walker (FRW) universe is spatially closed,
then the Einstein equation
\begin{equation}
H^2 = \kappa\rho -\frac{k}{a^2}\, , \qquad \kappa = \frac{8\pi G}{3}~, \label{eq:FRW}
\end{equation}
with $k = 1$ leads to the following cyclic solution for pressureless matter
\begin{equation}
a(\eta) = A(1 - \cos \eta) \, , \qquad t = A(\eta -\sin \eta) \, . \label{eq:cycloid}
\end{equation}
where $\eta = \int dt/a(t)$.

\begin{figure}[htp]
\centerline{\includegraphics[width=.5\textwidth]{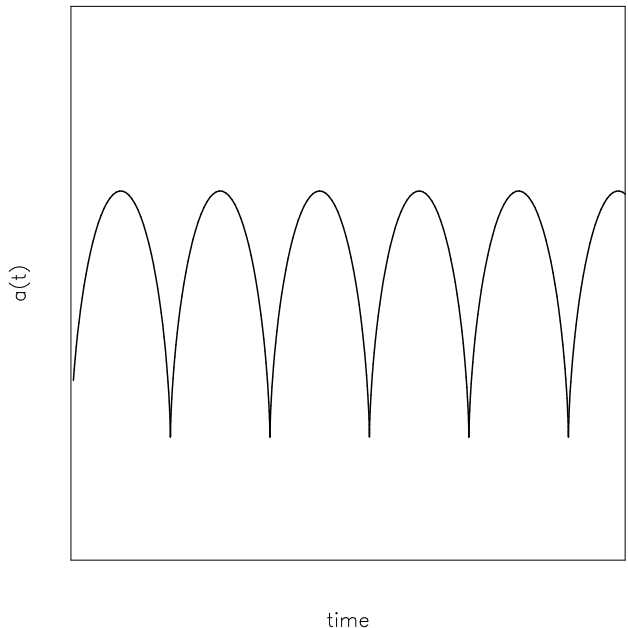}}
\caption{The expansion factor of a spatially closed matter-dominated universe
is described by the self-similar cycloid shown above. \label{fig:cycloid}}
\end{figure}

Solution (\ref{eq:cycloid}), depicted in figure \ref{fig:cycloid}, has an initial
singularity at $a = 0$, which is periodically repeated. This singularity is absent in
several modified-gravity theories. For instance, in a class of braneworld models
\cite{yuri,Maeda:2007cb} and also in loop quantum cosmology (LQC) \cite{Ashtekar:2006uz,
Calcagni:2004wu, Copeland:2005xs, Singh:2006sg, Stachowiak:2006uh}, one
gets\footnote{Braneworld and LQC corrections are not the only means of making the
universe bounce. A comprehensive review of bouncing cosmologies is provided in
\cite{brazil}.}
\begin{equation}
H^2 = \kappa\rho \left\lbrace 1 - \frac{\rho}{\rho_c}\right\rbrace -\frac{k}{a^2} \, ,
\qquad \kappa = \frac{8\pi G}{3} \, , \label{eq:bounce}
\end{equation}
which ensures that the universe `{\em bounces\/}' when $\rho \simeq \rho_c$. At late
times, when $\rho \ll \rho_c$, equation (\ref{eq:bounce}) reduces to (\ref{eq:FRW}). The
value of $\rho_c$ is related to the fundamental parameters appearing in braneworld
cosmology/LQC. We do not write them explicitly since the precise form of $\rho_c$ will
not be required in this paper.

In an attempt to describe a cyclic universe with {\em progressively larger\/} cycles,
Tolman \cite{tolman} assumed the presence of a viscous fluid with pressure
\begin{equation}
p = p_0 - 3\zeta H \, , \label{eq:4}
\end{equation}
where $p_0$ is the equilibrium pressure, $H = {\dot{a}}/{a}$, and $\zeta$ is the
coefficient of bulk viscosity. From equation (\ref{eq:4}) one finds that $p < p_0$ during
expansion ($H > 0$), whereas $p > p_0$ during contraction ($H < 0$). This asymmetry
during the expanding and contracting phases results in the growth of both energy and
entropy. The increase in entropy makes the amplitude of successive expansion cycles
larger, leading to a progressively larger and older universe \cite{tolman}.

Tolman believed that thermodynamically recycling the universe would have a ``liberalizing
action on our general thermodynamic thinking'' since it would dispel the notion that
``the principles of thermodynamics necessarily require a universe which was created at a
finite time in the past and which is fated for stagnation and death in the future''
\cite{tolman}. Thus Tolman's oscillating universe presented an alternative to the idea of
the thermodynamic heat death postulated by nineteenth century physicists and popular in
the twentieth century as well.

Although Tolman linked the asymmetry in pressure during expansion and collapse to
dissipation via (\ref{eq:4}), such an asymmetry arises even for non-dissipative systems
such as a massive scalar field in an FRW space-time.

\begin{figure}[htp]
\centerline{\includegraphics[width=.75\textwidth]{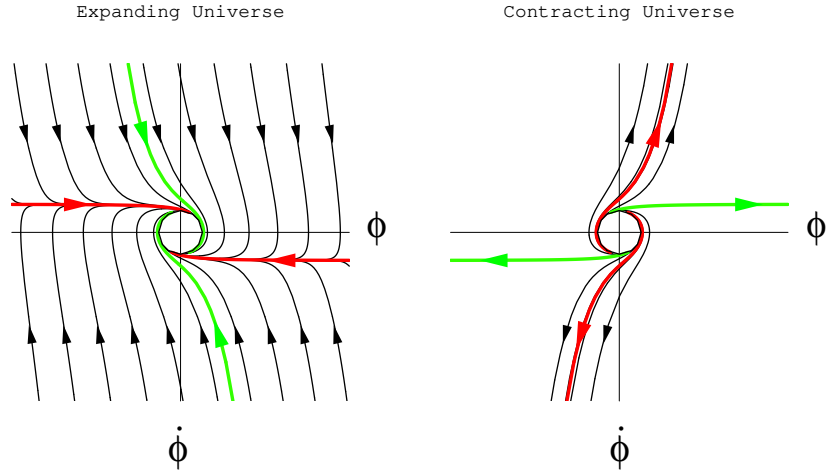}}
\caption{The attractor during expansion is ${\dot\phi}^2 \ll V \Rightarrow p \simeq -
\rho$, whereas during contraction the attractor is ${\dot\phi}^2 \gg V \Rightarrow p
\simeq \rho$; from \cite{kofman} (figure Copyright 2002 by the American
Physical Society). In a cyclic universe, these two attractor regimes give rise to {\em
cosmological hysteresis\/} which is shown in figure \ref{fig:hysteresis}.
\label{fig:phase_space}}
\end{figure}

\begin{figure}[htp]
\centerline{\includegraphics[width=.5\textwidth]{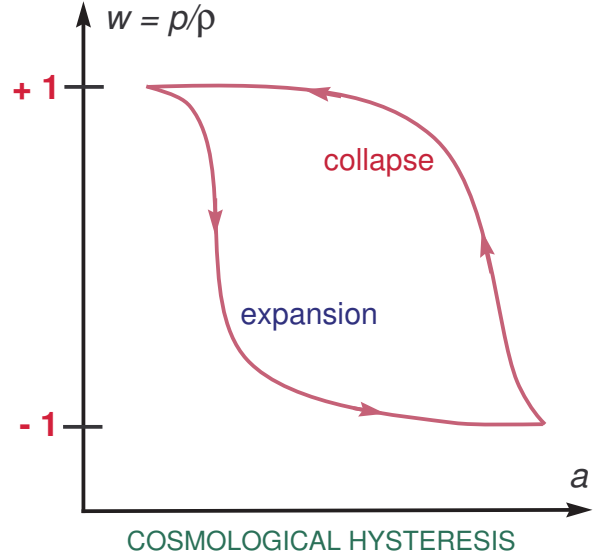}}
\caption{The {\em hysteresis loop} has $\oint p dV < 0$ evaluated over a single
expansion-contraction cycle in a scalar field dominated universe \cite{lesha}
(figure Copyright 2012 by the American Physical Society). \label{fig:hysteresis}}
\end{figure}

The energy density and pressure of a homogeneous scalar field are, respectively,
\begin{equation}
\rho = \frac12 \dot{\phi}^2 + V (\phi) \, , \qquad  p = \frac12 \dot{\phi}^2 - V (\phi)
\, , \label{eq:scalar_rho}
\end{equation}
and the scalar field equation of motion is
\begin{equation}
\ddot \phi + 3 H \dot \phi + \frac{dV}{d\phi} = 0 \, . \label{eq:scalar field}
\end{equation}
The term $3 H \dot \phi$ in (\ref{eq:scalar field}) acts like friction and damps the
motion of the scalar field when the universe expands ($H>0$). By contrast, in a
contracting ($H<0$) universe, the term $3 H \dot \phi$ acts like {\em anti-friction\/}
and accelerates the motion of the scalar field. Consequently, a scalar field with the
potential $V = m^2\phi^{2}$ displays two attractor regimes
\cite{belinsky1,belinsky2,belinsky3,kofman}
\begin{equation} \label{eq:8}
\begin{array}{lll}
p \simeq - \rho \quad &\mbox{during expansion} \quad &(H > 0) \, , \\
p \simeq \rho \quad &\mbox{during contraction} \quad &(H < 0) \, ,
\end{array}
\end{equation}
which are depicted in figure \ref{fig:phase_space}. In cyclic cosmology, these two
regimes combine together to produce {\em cosmological hysteresis\/} \cite{kanekar,lesha},
which is shown in figure \ref{fig:hysteresis}.

\begin{figure}[htp]
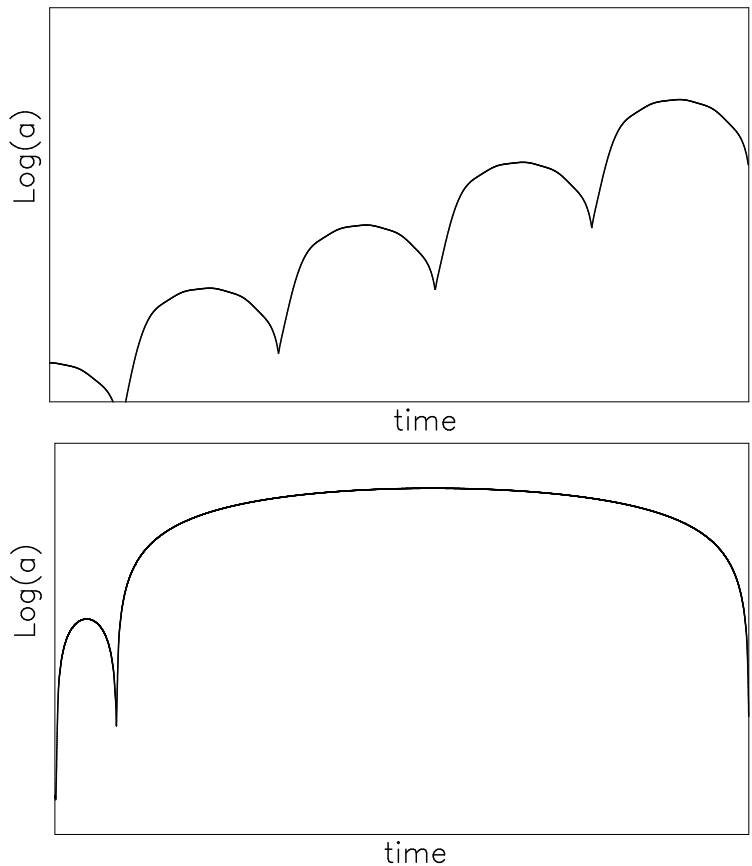

\centerline{\includegraphics[width=.6\textwidth]{phi-4a.ps}} \smallskip
\centerline{\includegraphics[width=.6\textwidth]{phi-4b.ps}}
\caption{The presence of hysteresis, shown in figure \ref{fig:hysteresis}, {\em
increases\/} the amplitude of successive expansion maxima and endows a cyclic universe
with an {\em arrow of time\/}. In the upper panel, cosmic turnaround is caused by the
presence of a {\em negative\/} cosmological constant
and the relevant expression linking the increase in successive expansion cycles
to hysteresis is (\ref{eq:hyst2}).
In the lower panel,
turnaround is due to the
 (positive) curvature term in (\ref{eq:FRW}).
In this case the expression describing increasing expansion cycles is (\ref{eq:hyst1}).
In both panels the cyclic
universe is sourced by a homogeneous scalar field with the potential $V(\phi) =
\frac{1}{2}m^2\phi^2$; also see \cite{lesha}. }
\label{fig:phi2}
\end{figure}

Cosmological hysteresis is a remarkable feature\footnote{Although this letter focuses on
the $m^2\phi^2$ potential, the phenomenon of hysteresis is quite general and exists for
other potentials as well, as demonstrated in \cite{kanekar,lesha}.} which the scalar
field shares with a viscous fluid by virtue of the parallel between (\ref{eq:8}) and
(\ref{eq:4}). The system of equations (\ref{eq:bounce}), (\ref{eq:scalar_rho}),
(\ref{eq:scalar field}) was numerically solved in \cite{kanekar,lesha}, and the results
are summarized in figure \ref{fig:phi2}. One finds that the presence of hysteresis leads
to local {\em time asymmetry\/} in cyclic cosmology, which is reflected in the increase
in amplitude of successive expansion cycles. This increase is related to the amount of
hysteresis in a given cycle, namely $\oint pdV$. If cosmic contraction is sourced by the
$k/a^2$ term in (\ref{eq:bounce}), then the increase in successive expansion maxima is
given by  \cite{kanekar,lesha}
\begin{equation}
\delta a_{\rm max} = - \kappa \oint pdV ~, \label{eq:hyst1}
\end{equation}
where $\oint pdV < 0$. Because of the relation $\Omega - 1 = (aH)^{-2}$, the phenomenon
of hysteresis leads to a universe which progressively becomes older, larger and more
spatially flat with each successive cycle. In addition, cosmological hysteresis ensures
that, for a potential such as $m^2 \phi^2$, the inflaton field $\phi$ gets driven to
larger values at the commencement of each new cycle. This ensures that the universe will
ultimately inflate, even if it did not do so initially. Thus inflation can commence from
a larger class of initial conditions in the presence of hysteresis than in its absence
\cite{kanekar,lidsey}.

Note that the universe can turnaround and contract even {\em in the absence of the
curvature term\/} provided its energy density can assume negative values. This can be
accomplished in a number of ways: (i)~Scalar-field potentials such as $V(\phi) =
\lambda\phi^4-m^2\phi^2$ and $V(\phi) \propto \cos{\phi}$ permit $V(\phi)$ to evolve to
negative values as the universe expands, leading eventually to contraction. (ii)~A
phantom field ($w < -1$) can give rise to a cyclic universe \cite{freese} within the
framework of the braneworld/LQC equation (\ref{eq:bounce}). (iii)~The presence of a small
{\em negative\/} cosmological constant, $\Lambda<0$, can make the universe turn around
and contract, giving rise to cyclicity. In this case, the field equations become
\begin{equation}
H^2 = \kappa\rho\left\lbrace 1 - \frac{\rho}{\rho_c}\right\rbrace -\frac{\vert\Lambda\vert}{3}\, ,
\qquad \kappa = \frac{8\pi G}{3} \, , \label{eq:bounce1}
\end{equation}
and the increase in successive expansion maxima is related to the quantum of hysteresis
by
\begin{equation}
\delta a^3_{\rm max} = \frac{\kappa}{\Lambda} \oint p d V \, . \label{eq:hyst2}
\end{equation}

The hysteresis equations (\ref{eq:hyst1}), (\ref{eq:hyst2}) are quite robust, and follow
simply from the relationship $\delta M + p\delta V = 0$ (where $M = \rho a^3$), which is
easily derived from the conservation condition $\nabla_a T^a{}_b =0 \Rightarrow
{\dot\rho} + 3H(\rho+p)=0$.

\begin{figure}[htp]
\centerline{\includegraphics[width=.9\textwidth]{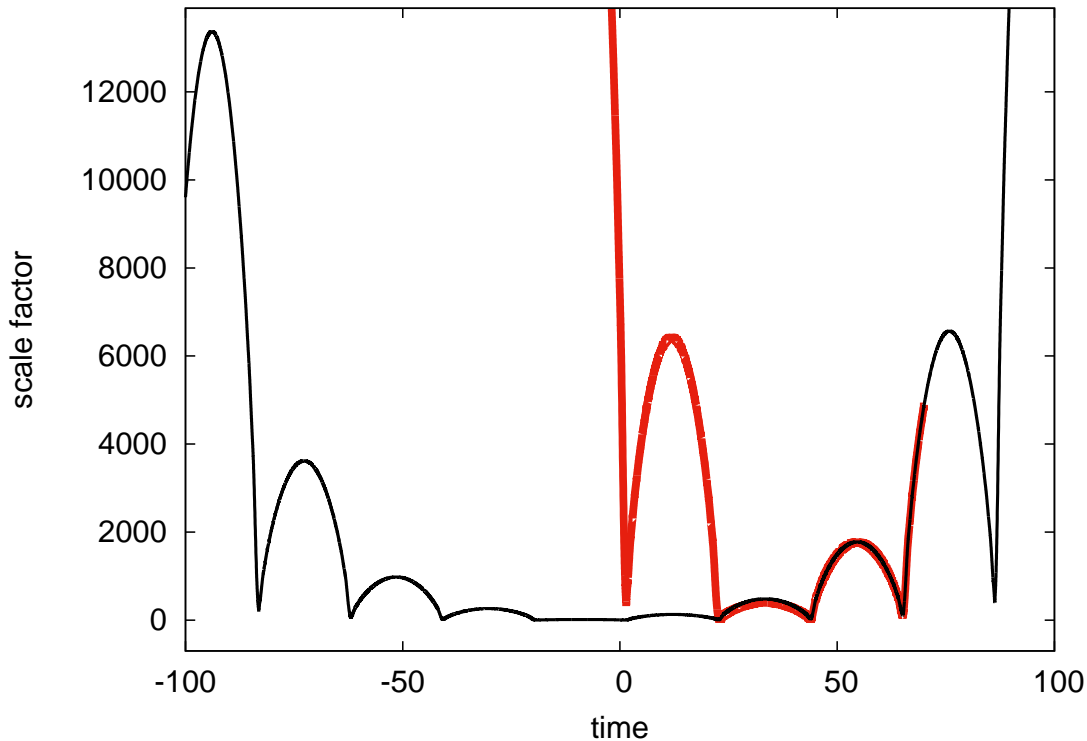}}
\caption{The expansion factor of a cyclic universe filled with a massive scalar field and
evolved according to (\ref{eq:bounce}), (\ref{eq:scalar_rho}), (\ref{eq:scalar field}).
The black (thin) curve is obtained by specifying {\em a priori\/} initial conditions at
$t = 0$ and integrating forward and backward in time. One finds that the expansion factor
evolves to larger values in both time directions indicating that cosmological hysteresis
is {\em incremental} ($\delta a_{\rm max} > 0$) in both cases. The red (thick) curve is
obtained by adding a small amount of noise (at the relative level of $10^{-6}$) to the
phase space variables at $t = 70$ and integrating backwards in time from this point. In
this case the trajectory retraces its steps from $t=70$ to $t \simeq 20$ displaying
decremental hysteresis ($\delta a_{\rm max} < 0$) during this period. However after the
first few cycles the expansion factor rapidly deviates from its original form and starts
showing incremental hysteresis ($\delta a_{\rm max} > 0$). The behaviour of the red
(thick) curve thus illustrates that the scalar-field dominated universe with decremental
hysteresis is {\em unstable\/} to small perturbations.  In this simulation, we express
all dimensional variables using reduced Planck units with mass scale $M_* = \sqrt{3/ 8
\pi} M_{\rm P}$, where $M_{\rm P} = 1 / \sqrt{G}$ is the Planck mass, and we have chosen
$m = 3.5$, $\rho_c = 60$, which corresponds to the physical parameters $m \approx 1.2\,
M_{\rm P}$ and $\rho_c \approx 0.85\, M_{\rm P}^4$.  The initial conditions are chosen to
be $\phi = 0.1$, $\dot \phi = - 0.5$, and $a = 5$.   We should add, however, that the
qualitative behaviour of the expansion factor shown above is generic and is common to a
large class of initial conditions.
\label{fig:u-shape}}
\end{figure}

Consider next an arbitrary trajectory $a (t)$, $\phi (t)$ of the cosmological system
described by (\ref{eq:bounce}), (\ref{eq:scalar_rho}), (\ref{eq:scalar field}) with $k =
1$. Since the scale factor is bounded from below by $a^2 \geq 4/\kappa \rho_c$, there is,
typically, a point on this trajectory where it reaches the minimal value $a = a_{\rm min}
> 0$. In our simulations, we always observe one such point at which the universe attains
its smallest size.  The neighborhood of this point can then be viewed as the ``origin''
of the arrow of time for this particular evolution.  From this moment of time, the
universe evolves with hysteresis and with systematically increasing scale factor in both
time directions. We illustrate this U-shaped behaviour by the black (thin) curve in
figure \ref{fig:u-shape}.

The above description suggests that the scale factor provides us with a natural arrow of
time. The reason behind this lies in the fact that $a$ is the only variable in our phase
space that is unbounded and unconstrained from above.  In contrast, both the scalar field
$\phi$ and its time derivative $\dot \phi$ are bounded by the condition $\rho \leq
\rho_c$ valid for our system (\ref{eq:bounce}). The three-dimensional phase-space of our
system, $\lbrace a,\phi,{\dot\phi}\rbrace$, can therefore  be compared to a particle
moving in a semi-infinite curved pipe whose one end is blocked by a wall. Bouncing back
and forth off the walls of the pipe, the particle will typically move further and further
away from the wall. Likewise, our system {\em typically\/} escapes to $a \to \infty$ as
time goes on (in both time directions), under the joint influence of cyclicity and
hysteresis.

It is interesting that the `one-past--two-futures' behaviour encountered here has also
been observed for an $N$-body system consisting of dissipationless particles interacting
via Newtonian gravity \cite{Barbour:2013jya,Barbour:2014bga}. We discuss some
similarities between our system and gravitational clustering in
section~\ref{sec:discussion}.

\subsection*{Issues relating to the stability of trajectories}

To obtain a U-shaped trajectory such as the black (thin) curve in figure
\ref{fig:u-shape}, we start with some initial conditions at $t = 0$ and evolve the system
both to the past and to the future of this moment. In all cases that we have studied,
with {\em a priori\/} initial conditions at $t = 0$, we observe that the point of minimal
value of the scale factor lies quite close to the moment $t = 0$, and {\em incremental}
hysteresis ($\delta a_{\rm max} > 0$) quickly develops (after one or two cycles) in both
time directions.  In other words, by choosing the initial conditions at random, it
appears to be quite improbable to fall on a trajectory that displays {\em decremental\/}
hysteresis ($\delta a_{\rm max} < 0$) for a considerable number of cycles. The only way
to actually produce such a behavior is to time-reverse the trajectory with the usual {\em
incremental\/} hysteresis ($\delta a_{\rm max} > 0$). Thus, as we run the simulation
shown by the black (thin) curve in figure \ref{fig:u-shape} from $t = 70$ backwards in
time with the time-reversed initial conditions at $t = 70$, we follow exactly the same
evolution in reverse order, which, therefore, initially displays decremental hysteresis.

However, it is interesting to note that evolution in the direction of decremental
hysteresis is unstable with respect to small variations in the initial conditions.  We
demonstrate this by perturbing the values of the phase-space variables at $t = 70$ at the
small relative level of $10^{-6}$ and running the simulation backwards in time from this
point.  In this case, we follow the trajectory shown by the red (thick) curve in figure
\ref{fig:u-shape}. After a couple of cycles, this (red) trajectory begins to deviate from
its original (black) trajectory and soon enters the regime of incremental hysteresis.  In
marked contrast to this behaviour, evolution in the direction of incremental hysteresis,
when subject to similar perturbations, demonstrates perfect stability.

All this appears to suggest that we may be witnessing a phenomenon which, in some sense,
is ``physically'' irreversible, meaning that reversibility is lost in the presence of
even tiny perturbations. (This is similar to the process of statistical equilibration in
a mechanical system with many degrees of freedom.) One may wonder about the reasons
behind the time irreversibility displayed by our system. We believe that it is related to
the different nature of its attractors during expansion and contraction.  We have seen
earlier that, for scalar-field driven cosmology, the equation of state of an attractor
during expansion, $p \simeq -\rho$, is replaced by $p \simeq \rho$ during contraction
(see figure \ref{fig:phase_space}).  The stable inflationary regimes during expansion
lead to an ever increasing size of the universe with each cycle, and therefore to
incremental hysteresis. On the contrary, deflationary regimes during contraction (which
would be responsible for decremental hysteresis) are highly unstable. Hence, they are
improbable, and this is the reason why it is so unlikely to observe decremental
hysteresis with {\em a priori\/} initial conditions. We stress here that the {\em time
irreversibility\/} which we have encountered is manifest on sufficiently long timescales,
involving several expansion-contraction cycles of the universe.\footnote{While increasing
expansion cycles are routinely encountered for flat potentials, steep potentials can give
rise to a two-fold cyclic pattern in time with smaller expansion cycles being nested in
larger ones, much like the phenomenon of `beats' in acoustics \cite{lesha}.}

\section{Inflation}
\label{sec:inflation}

As demonstrated above, the cyclic universe appears to possess an arrow of time which may
be related to the fact that stable trajectories become unstable under time reversal.
Another example of a dissipationless system with similar properties (with respect to time
reversal) is provided by inflation. The equation of motion (\ref{eq:scalar field}) of the
inflaton field is very similar to the motion of a body falling under gravity in a viscous
medium characterized by a viscosity coefficient $\eta$:
\begin{equation} \label{viscous}
m {\ddot z} + \eta {\dot z} - mg = 0 \, .
\end{equation}
In both cases, an attractor is soon reached. In the case of (\ref{viscous}), this
attractor is the terminal velocity $v \equiv {\dot z} = mg/\eta$. It is reached by means
of the transfer of translational energy of the body to the large number of translational
and rotational degrees of freedom present in (molecules of) the viscous medium. This
transfer results in an entropy increase, which converts the free-fall of the body into an
irreversible process.

\begin{figure}[htp]
\includegraphics[width=.49\textwidth]{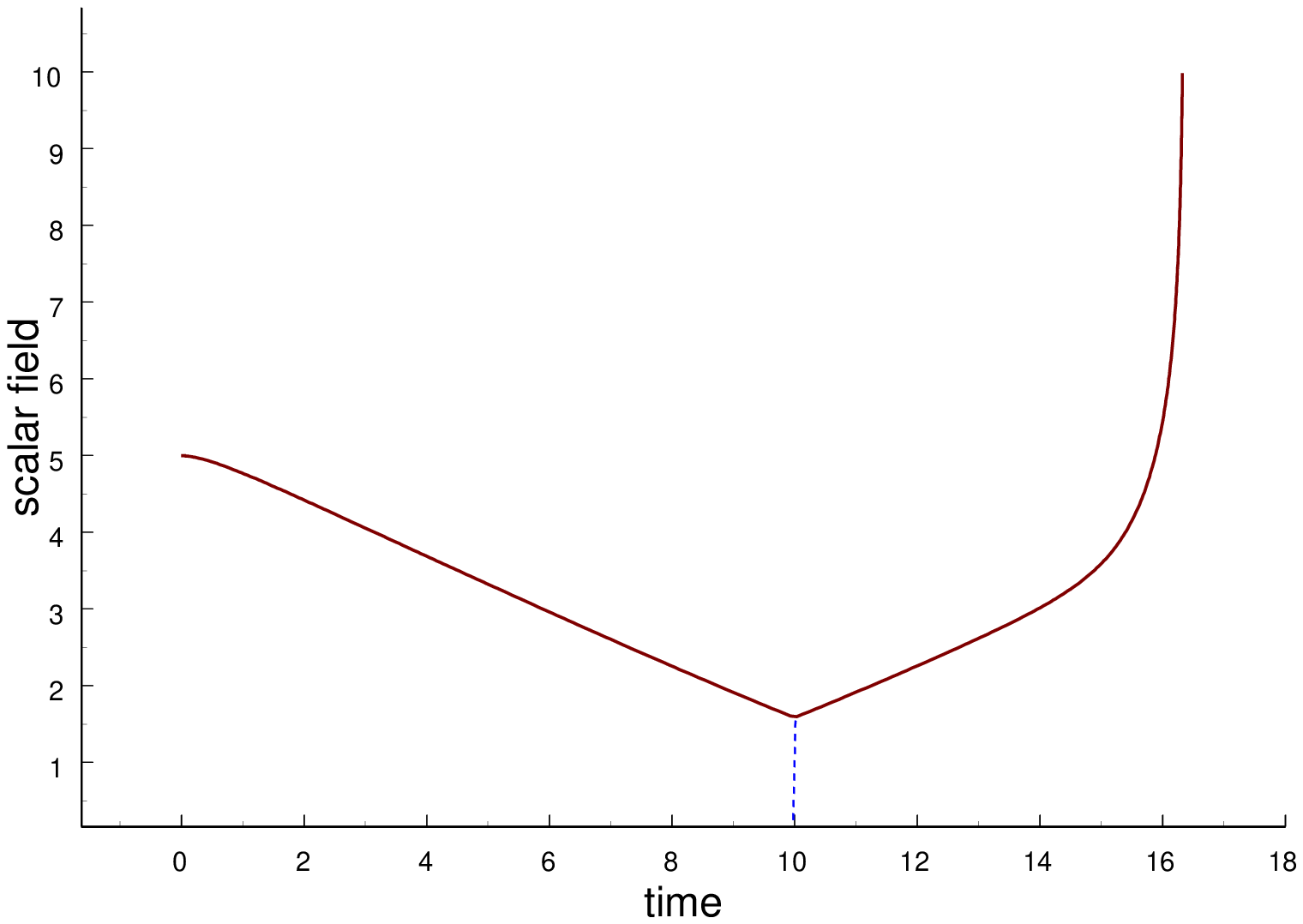}
\includegraphics[width=.49\textwidth]{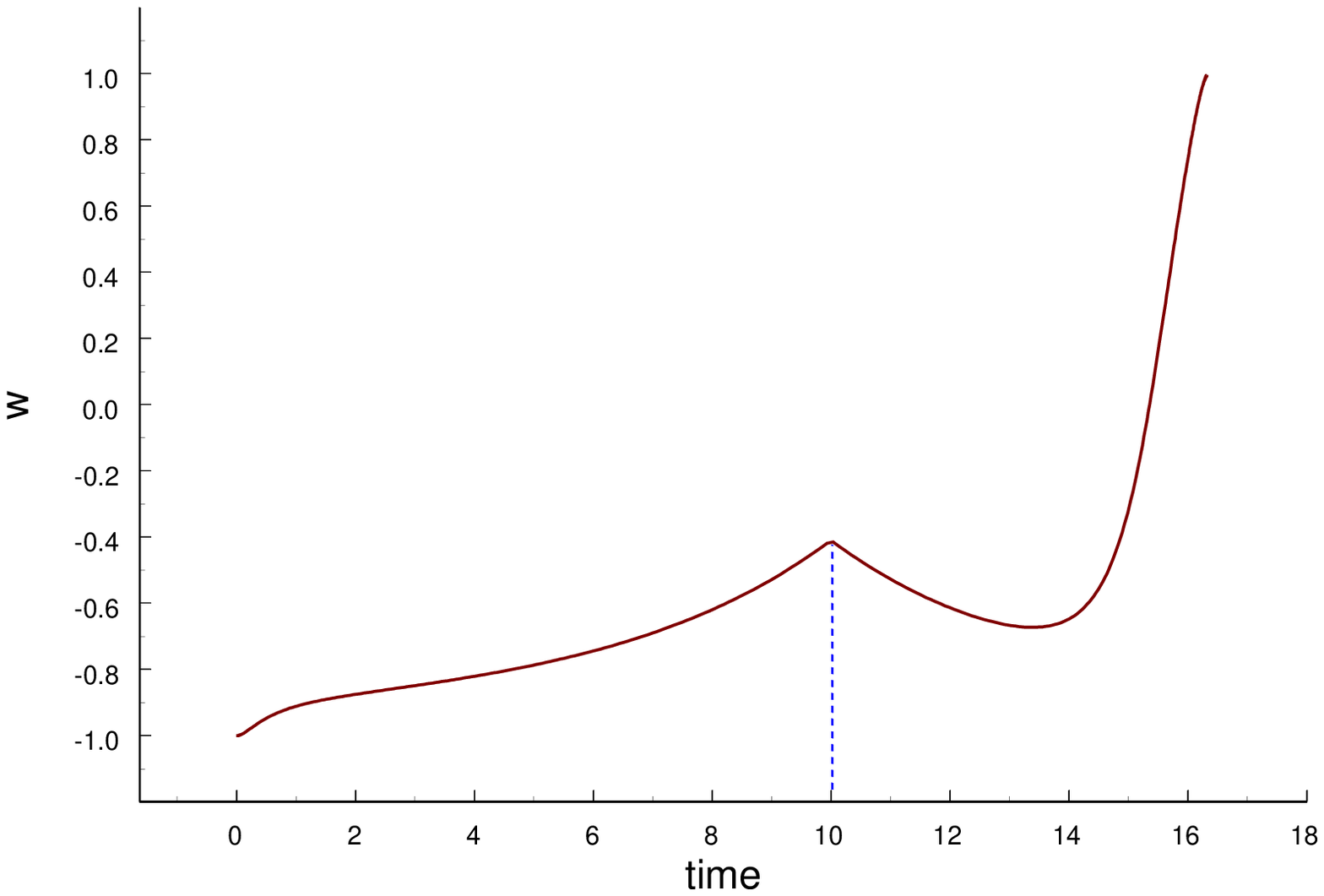}
\includegraphics[width=.49\textwidth]{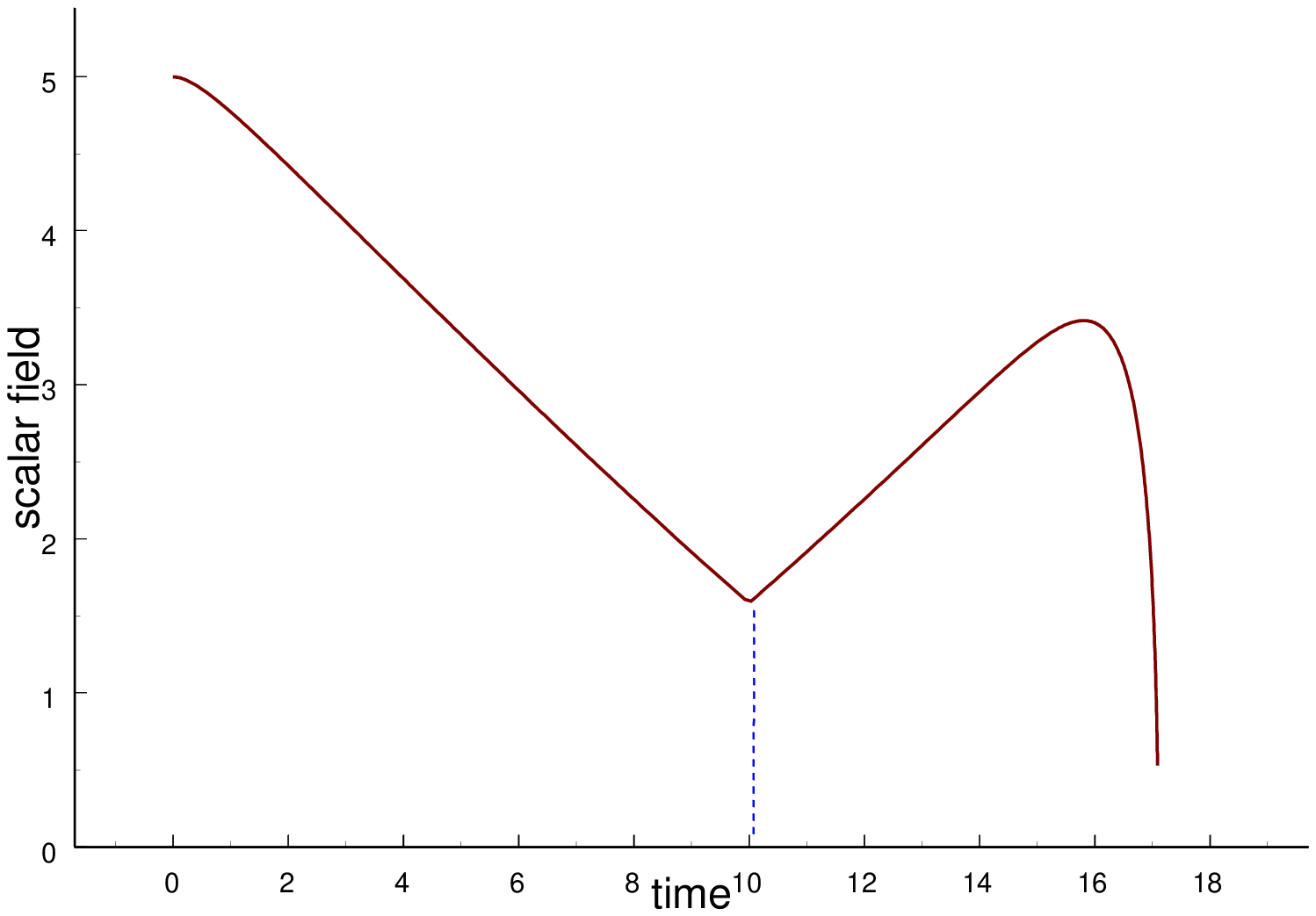}
\includegraphics[width=.49\textwidth]{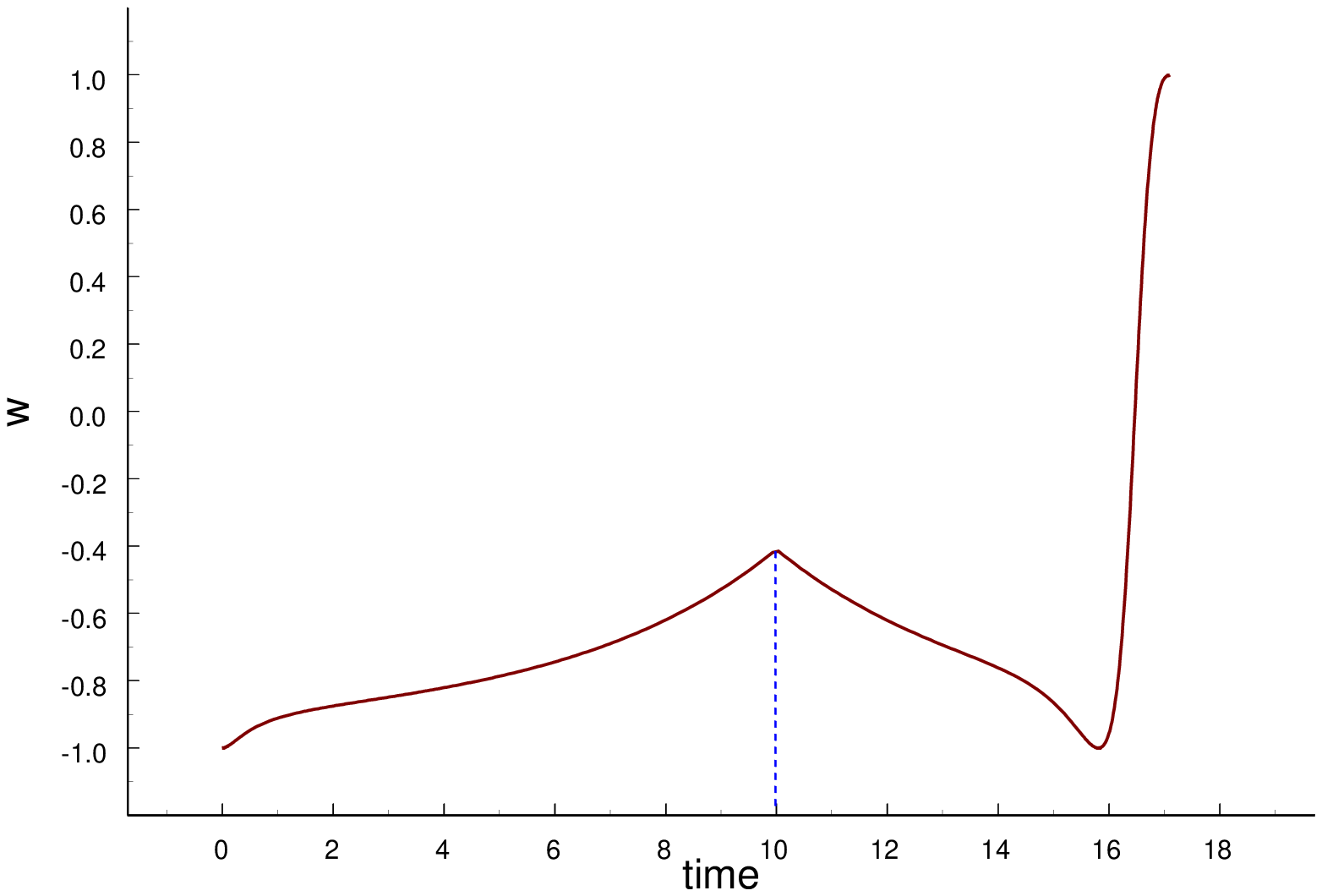}
\caption{Scalar-field values (left) and the associated equation of state $w=p/\rho$
(right) for a universe undergoing inflation via the potential $V = m^2\phi^2/2$. (Units
are in terms of the Planck scale.) The trajectory of $a(t)$, $\phi(t)$ is reversed at $t
\simeq 10$ via ${\dot\phi} \to - {\dot\phi}$, $H \to -H$ (vertical blue line) by
simultaneously perturbing $H$ by $10^{-4}$ (upper panel) and $10^{-6}$ (lower panel) of
its original value. Since the equations of motion are time-reversible, one might expect the
time-reversed trajectory at $t > 10$ to be a mirror image of the forward-time trajectory
at $t < 10$. Surprisingly, this is {\em not the case}. For a while after time reversal
($10\leq t \leq 14$), $\phi(t)$ does follow its forward time evolution `backward in
time', with $w$ {\em decreasing\/} as expected. But for $t > 14$ the time-reversed
trajectory becomes highly unstable causing $w$ to increase towards $+1$ instead of
decreasing to $-1$.  \label{fig:inflation}}
\end{figure}

Similarly, a scalar field rolling down a potential with a gentle slope soon reaches the
slow-roll limit: $|{\dot\phi}| \simeq |V'/3H|$, $| {\ddot \phi} | \ll |3 H \dot \phi |$.
In this case, the role of viscosity is played by the expansion of the universe, so that
$\eta \equiv 3H$.  An important difference from (\ref{viscous}) is that, since equation
(\ref{eq:scalar field}) is not associated with microscopic entropy production, the
universe precisely follows its time-reversed trajectory after the signs of $H$ and
${\dot\phi}$ have been reversed.  However, after time reversal, the Hubble parameter
changes sign, which makes the second term in (\ref{eq:scalar field}), which originally
acted like friction, to behave like anti-friction, rendering the time-reversed slow-roll
trajectory {\em unstable\/} to small perturbations. Since small perturbations, in the
form of quantum fluctuations, are always present in the inflationary scenario, one comes
to the conclusion that, practically speaking, an inflationary trajectory is difficult to
reverse, as illustrated in figure \ref{fig:inflation}. As shown in that figure, a larger
perturbation causes, upon time reversal, an earlier departure from the inflationary
equation of state $w \simeq -1$. Moreover, one sees that the perturbation in the left
panel of figure \ref{fig:inflation} (in which the Hubble parameter has been perturbed by
$10^{-4}$ times its original value) leads to the asymptotic behaviour $\phi \to +\infty$,
whereas perturbing the time derivative of the scale factor by $10^{-6}$ of its original
value induces the $\phi \to -\infty$ regime. In a more realistic scenario involving
inhomogeneous perturbations, a rapid deviation from the inflationary regime would be
accompanied by growth in scalar-field inhomogeneity.  This issue, however, lies beyond
the scope of the present paper.

The departure from the inflationary regime during contraction leads to a larger value of
the scale factor as the Planck density is reached. Consequently, in the presence of a
hypothetical Planck-scale bounce, the next cycle commences from a larger value of the
scale factor than the previous one, as shown in figure \ref{fig:u-shape}. (This happens
because the energy density, $\rho_\phi$, varies slowly near the de~Sitter regime but
grows much more rapidly: as $a^{-6}$, during the asymptotic $w \simeq 1$ stage of
expansion displayed in the right panel of figure \ref{fig:inflation}).

The fact that the inflationary trajectory is difficult to reverse has important
consequences.

It is well known that the homogeneous and isotropic FRW universe, in which matter
satisfies the `energy conditions', comprises a set of measure zero in the space of all
homogeneous but anisotropic solutions of the Einstein equations \cite{ch73}. Taken at
face value, this might suggest that the universe unravelled from a very special set of
initial conditions. However, the inflationary paradigm throws a fresh perspective on this
issue since, an anisotropic universe containing a positive cosmological constant rapidly
isotropises, approaching the de Sitter space attractor if the spatial curvature is not
too large \cite{no-hair}. This remains true for inflationary space-times in which a
scalar field replaces $\Lambda$ \cite{ms86,no-hair1}. In this case, the de Sitter space
attractor solution is a transient, and lasts while the slow roll conditions are
satisfied. Consider, for instance, the Bianchi I model
\begin{equation}
ds^2 = dt^2 - \sum_{i=1}^3a_i^2(t)(dx^i)^2 \, ,
\end{equation}
for which the equations of motion become
\begin{eqnarray}
&&H^2 = \frac{8\pi G}{3}\rho + \frac{1}{2}{\dot\phi}^2 + V(\phi) + \rho_{\rm anis}~,
\label{eq:hubble_bianchi}\\
&&\ddot \phi + 3 H \dot \phi + \frac{dV}{d\phi} = 0~, \label{eq:field_bianchi}
\end{eqnarray}
where $H$ describes the mean expansion rate
\begin{equation}
H = \frac{1}{3}\sum_{i=1}^3\left (\frac{{\dot a_i}}{a_i}\right ) = \frac{\dot a}{a}~,
\qquad a = \left (a_1a_2a_3\right )^{1/3}~,
\end{equation}
and the anisotropy is encoded in the anisotropy density $\rho_{\rm anis}$\,:
\begin{equation}
\rho_{\rm anis} = \frac{1}{6}\sum_{i=1}^3\left (\frac{{\dot a_i}}{a_i} - H\right )^2 =
\frac{\Sigma}{a^6}~.
\end{equation}

From (\ref{eq:hubble_bianchi}) we see that a large initial anisotropy, $\rho_{\rm anis}$,
increases the expansion rate, $H$, thereby introducing greater damping into the scalar
field equation of motion (\ref{eq:field_bianchi}). The inflationary attractor $P \simeq
-\rho$ in (\ref{eq:8}) can therefore be more rapidly reached in the presence of
anisotropy than in its absence \cite{ms86,maartens}. The observation that inhomogeneities
and anisotropies get rapidly diluted in an inflationary universe has occasionally been
referred to as the `cosmic no-hair theorem' in analogy with no-hair theorems for black
holes proven during the 1970's.

However, the fact that the equations governing inflationary expansion are reversible has
sometimes been used to undermine the no-hair theorem by means of the following argument.
Let us consider a hypothetical universe, far more anisotropic than ours, in which the
present anisotropy parameter is $\rho^{\rm final}_{\rm anis}$. Then, since the equations
of motion (\ref{eq:hubble_bianchi}), (\ref{eq:field_bianchi}) are reversible, one can
evolve them back in time in order to determine the initial anisotropy,  $\rho^{\rm
initial}_{\rm anis}$, which would have led to $\rho^{\rm final}_{\rm anis}$ after
inflation. In this manner, it is always ``possible to choose initial data in such a way
that the universe would be more anisotropic than ours is today'' \cite{ms86}. The
weakness in this argument is connected with the issue of the probability of such initial
conditions. In the previous section, we demonstrated that randomly chosen {\em a
priori\/} initial conditions always lead to incremental hysteresis, hence, develop an
arrow of time from the very beginning.  If the conditions leading to decremental
hysteresis are indeed quite improbable, then the presence of small perturbations (in the
form of vacuum fluctuations) makes the equations of motion of the scalar field
effectively non-reversible. This observation considerably weakens the above argument and
strengthens the no-hair theorem. (We propose to study this issue in greater detail in a
companion work.)

\section{Discussion}
\label{sec:discussion}

We have demonstrated the appearance of an arrow of time in the time-reversible
cosmological system consisting of a spatially closed universe filled with a homogeneous
scalar field. This situation is not uncommon, and takes place in statistical physics with
many degrees of freedom, where, for example, the (classical) molecular dynamics is
evidently reversible, but small deviations from the exact time-reversed velocities (which
would lead to an evolution with decreasing entropy) rapidly lead to the usual entropy
increase. In the case of cyclic cosmology, the presence of an arrow of time is intimately
linked to the phenomenon of hysteresis. Four important conditions need to be satisfied
for the occurrence of hysteresis:

\begin{enumerate}

\item The potential for $V(\phi)$ should allow the occurrence of inflation.

\item \label{item} $V(\phi)$ should have a minimum about which the scalar field
    oscillates after inflation. This condition is essential since it allows the field
    $\phi$ to {\rm phase-mix}. When this condition is relaxed, for instance in
    monotonically decreasing potentials such as $V(\phi) \propto \phi^\alpha,
    \,\alpha < 0$, hysteresis is absent and all cycles have the same amplitude
    \cite{lesha}.

\item A condition for turnaround is present, either in the form of positive spatial
    curvature or a negative cosmological constant, or a potential $V(\phi)$ which
    becomes negative at late times.

\item The initial big-bang singularity is replaced by a bounce at large
    densities.\footnote{It is well known that a bounce can be produced within GR in a
    model with positive spatial curvature, although this requires fine-tuning of
    initial conditions for the scalar field \cite{Starobinsky}. However, this bounce
    takes place when the equation of state parameter is close to $w = -1$ in {\it
    both\/} the contracting stage and the follow-on expanding stage. Since no
    hysteresis is present in this case, such a bounce cannot cause the effects
    studied in the present paper.}

\end{enumerate}

If the above conditions are met, then the FRW cosmology of a scalar field with {\em a
priori\/} initial conditions exhibits an arrow of time, with preferred evolution towards
larger scale factors from cycle to cycle. Phenomenologically, the dissipationless
universe with a scalar field behaves as if it contained a fluid with bulk viscosity, with
the asymmetry in pressure during expansion and contraction giving rise to the phenomenon
of cosmological hysteresis.\footnote{Within a multiverse setting, the presence of
hysteresis allows the universe to experience many different vacua during successive
cosmological cycles \cite{piao}.}

One should mention here that an arrow of time has also been discovered quite recently in
the very different context of Newtonian dynamics \cite{Barbour:2013jya,Barbour:2014bga}.
It was noticed that an $N$-body system of point masses with Newtonian gravitational
attraction can be endowed with a specific quantity, called {\em complexity\/} by the
authors, whose value reaches a minimum on every $N$-body trajectory with total
centre-of-mass energy and total angular momentum both equal to zero. Complexity is given
by the ratio $C_{\rm S} = \ell_{\rm rms} / \ell_{\rm mhl}$, where
\begin{equation} \label{rms}
\ell_{\rm rms} = \frac{1}{m_{\rm tot}} \left( \sum_{a < b} m_a m_b r_{ab}^2 \right)^{1/2}
= \left( \frac{I_{\rm cm}}{m_{\rm tot}} \right)^{1/2}
\end{equation}
is the root-mean-square length, which is expressed through the centre-of-mass moment of
inertia $I_{\rm cm}$, and the mean harmonic length $\ell_{\rm mhl}$ is defined by
\begin{equation} \label{mhl}
\frac{1}{\ell_{\rm mhl}} = \frac{1}{m_{\rm tot}^2} \sum_{a < b} \frac{m_a m_b}{ r_{ab}} \, .
\end{equation}
Complexity displays a U-shaped behaviour since its value systematically decreases before
the minimum point and systematically increases after it.  This quantity can thus play the
role of an entropy, and the point of its minimum can be regarded as a point where the
arrow of time changes direction.\footnote{Note that the centre-of-mass moment of inertia
on a trajectory with total centre-of-mass energy $E \geq 0$ satisfies $\ddot I_{\rm cm} >
0$, also reaching a minimum on every such trajectory. One should also note an important
difference between the usual entropy and `complexity'. Unlike entropy, the complexity
$C_{\rm S}$ is not additive for a system consisting of several `islands' separated by
large distances relative to their sizes. Indeed, while the inverse mean harmonic length
(\ref{mhl}) is approximately additive, the root-mean-square length (\ref{rms}) is far
from being an approximate constant under the addition of new islands to the system. Thus,
the moment of time where $C_{\rm S}$ reaches a minimum for the whole system consisting of
several remote islands may be quite different from the moments of time where this
quantity reaches minima for individual islands, or even from the moment when the sum of
$C_{\rm S}$ over the islands reaches a minimum. } We encountered this U-shaped
`one-past--two-futures' behaviour earlier, in section~\ref{sec:cycle}, in our discussion
of hysteresis. We elaborate on this issue once more, in view of its importance, since it
has appeared in two completely distinct discussions of the arrow of time.

The black (thin) curve in Figure~\ref{fig:u-shape} shows a typical trajectory in terms of
the scale factor. There is a point around $t = 0$ where the scale factor takes a minimal
value.

As we move from $t = 0$ to the past or to the future, the universe displays hysteresis in
both time directions.  However, when the trajectory is viewed in one time direction, say,
from left to right, it is unstable in the region $ t < 0$ (decremental hysteresis,
$\delta a_{\rm max} < 0$), while it is stable in the region $t > 0$ (incremental
hysteresis, $\delta a_{\rm max} > 0$).

There are several conclusions that one can derive from this picture.

\begin{itemize}

\item A typical evolution of the universe in our model possesses hysteresis and an
    ``origin of time'' (region around $t = 0$), from which the evolution is similar
    in both time directions.

\item The trajectories are stable in the direction of time which displays incremental
    hysteresis ($\delta a_{\rm max} > 0 \Rightarrow \oint pdV < 0$), and are unstable
    in the opposite direction.

\item All these features are quite similar to the behaviour of entropy.  If we
    prepare a complicated physical system (a gas in a box) randomly distributed and
    in a non-equilibrium state and run it in both time directions, we will observe an
    entropy increase in both time directions.  The system will tend to thermodynamic
    equilibrium in both time directions, and this convergence to equilibrium will be
    stable with respect to perturbations of initial conditions. However the
    time-reversed evolution in both cases will be highly unstable.

\end{itemize}
We have thus demonstrated that our cyclic model possesses the U-shaped
`one-past--two-futures' behaviour discovered recently for an $N$-body system. The
scalar-field based cyclic universe may therefore be regarded as the simplest known
example of a dissipationless system possessing an arrow of time.\footnote{In contrast to
an $N$-body system, our dynamical system possesses only one and a half degrees of freedom
--- described by the scale factor and the scalar field with a single constraint.}

One should also note the similarity of the eternal classical universe depicted by the
black (thin) curve in figure \ref{fig:u-shape} to the scenario of eternal inflationary
universe due to Carroll and Chen \cite{Carroll:2004pn}.  In the latter, it is a
spontaneous quantum fluctuation in the scalar field in a vacuum de~Sitter space that
plays the role of the `origin' of time, which then flows towards eternal inflation and
entropy growth in two opposite directions. In our model, the universe is driven by a
classical scalar field during all its history, and quantum fluctuations are not taken
into account.

Finally, we draw attention to the fact that we do not consider spatial inhomogeneities in
the present work. Introducing such inhomogeneities would mean increasing the number of
degrees of freedom and creating new kinds of instability.  The fact that inhomogeneities
are growing during cosmological evolution creates the usual arrow of time for gravitating
systems --- a self-gravitating system evolves from a homogeneous to a highly
nonhomogeneous distribution (in contrast to a molecular gas, which evolves towards a
homogeneous configuration). The well-known proposal by Penrose to use the Weyl tensor as
a measure of gravitational entropy is based on this property (see for instance
\cite{Clifton} for the current state of this proposal).

Our study demonstrates the existence of cosmological situations where the effective arrow
of time is not associated with the Weyl tensor.  This is so because this tensor
identically vanishes for the FRW metric which we consider. This leads to the conclusion
that not all effectively time-irreversible phenomena in self-gravitating systems can be
reduced to measures based on the Weyl tensor (and, more generally, that there exist
time-irreversible phenomena which are not associated with the growth of spatial
inhomogeneities).

\section*{Acknowledgments}

V~S and Y~S acknowledge support from the India--Ukraine Bilateral Scientific Cooperation
programme. The work of Y~S is supported in part by the State Foundation of Fundamental
Research of Ukraine. The work of A~T is supported by RFBR grant 14-02-00894, and
partially supported by the Russian Government Program of Competitive Growth of Kazan
Federal University. The authors acknowledge insightful comments made by an anonymous
referee which helped to improve the quality of this paper.

\end{document}